\RequirePackage{amsmath}
\documentclass[runningheads]{llncs}

\usepackage{color} 
\usepackage{tikz}
\usepackage[utf8x]{inputenc}
\usepackage{amsmath} 
\usepackage{comment}
\usepackage{booktabs} 
\usepackage{makecell}
\usepackage{multirow}
\usepackage{graphicx} 
\usepackage{comment} 
\usepackage{balance}
\usepackage{todonotes}
\usepackage{algorithm,float}
\usepackage{nicefrac}
\usepackage[noend]{algpseudocode}
\usepackage{tabu}
\usepackage{cite}
\usepackage{amssymb}
\usepackage{etoolbox}
\usepackage{rotating}
\usepackage{enumitem}
\usepackage{color,soul}

\usepackage{array}

\usepackage[caption=false]{subfig}

\newcommand{\modSimple}{\textbf{\texttt{BSL}}}
\newcommand{\modInterval}{\textbf{\texttt{INT}}}
\newcommand{\modWindow}{\textbf{\texttt{WIN}}}
\newcommand{\modEvo}{\textbf{\texttt{EVO}}}
\newcommand{\modCoverage}{\textbf{\texttt{COV}}}
\newcommand{\modCovWin}{\textbf{\texttt{CWI}}}
\newcommand{\modEvoCovWin}{\textbf{\texttt{ECW}}}
\newcommand{\eventGraph}{\textit{Event Graph}}
\newcommand{\eventGraphs}{\textit{Event Graphs}}
\newcommand{\dbpediaEventGraph}{\textit{DBpedia Event Graph}}
\newcommand{\wikidataEventGraph}{\textit{Wikidata Event Graph}}
\newcommand{\knowledgeGraph}{\textit{Knowledge Graph}}
\newcommand{\textualFeatures}{\texttt{TEX}}
\newcommand{\spatioTemporalFeatures}{\texttt{STP}}
\newcommand{\embeddingFeatures}{\texttt{EMB}}
\newcommand{\logReg}{\texttt{LOG}}
\newcommand{\svm}{\texttt{SVM}}
\newcommand{\randomForest}{\texttt{RF}}
\newcommand{\stranse}{\texttt{STransE}}

\makeatletter
\renewcommand\section{\@startsection{section}{1}{\z@}%
                       {-5\p@ \@plus -4\p@ \@minus -4\p@}%
                       {6\p@ \@plus 4\p@ \@minus 4\p@}%
                       {\normalfont\large\bfseries\boldmath
                        \rightskip=\z@ \@plus 8em\pretolerance=10000 }}
\renewcommand\subsection{\@startsection{subsection}{2}{\z@}%
                       {-5\p@ \@plus -4\p@ \@minus -4\p@}%
                       {6\p@ \@plus 4\p@ \@minus 4\p@}%
                       {\normalfont\normalsize\bfseries\boldmath
                        \rightskip=\z@ \@plus 8em\pretolerance=10000 }}
\renewcommand\subsubsection{\@startsection{subsubsection}{3}{\z@}%
                       {-3\p@ \@plus -4\p@ \@minus -4\p@}%
                       {-1.5em \@plus -0.22em \@minus -0.1em}%
                       {\normalfont\normalsize\bfseries\boldmath}}
\makeatother

\setlist[itemize]{leftmargin=*, topsep=0pt}

\newcommand{\approach}{\texttt{Hap\-Pen\-Ing}}

\newcommand{\definitionskip}{\vspace{-4pt}}

\begin{document}

\hyphenation{dis-co-ve-ry Event-KG ana-ly-tics po-pu-la-ri-ty know-led-ge re-fe-ren-ce fle-xib-le se-cond he-te-ro-ge-neous se-ve-ral existen-ce fa-ci-li-tate cha-ra-cte-ris-tics pro-per-ty ori-gi-na-te de-ve-lo-ped re-le-van-ce Wi-ki-pe-dia ap-proach-es ap-proach-es
pro-per-ties ma-nual-ly ex-pe-ri-men-tal ge-ne-ra-ted mo-de-led mi-ni-mum ma-xi-mum ge-ne-ra-te ge-ne-ra-ting ge-ne-ra-ted co-ve-ra-ge equi-va-lent ave-ra-ge
in-di-vi-du-al other-wi-se cha-rac-teris-tic si-mi-la-ri-ty par-ti-cu-lar}

\author{Simon Gottschalk\orcidID{0000-0003-2576-4640} \and \\ Elena Demidova\orcidID{0000-0002-5134-9072}}

\institute{L3S Research Center, Leibniz Universität Hannover, Hannover, Germany \\
 \email{\{gottschalk, demidova\}@L3S.de}
} 

\title{HapPenIng: Happen, Predict, Infer --- \\ Event Series Completion in a Knowledge Graph}
\titlerunning{HapPenIng: Happen, Predict, Infer}

\maketitle

\begin{abstract}
Event series, such as the Wimbledon Championships and the US presidential elections, represent important happenings in key societal areas including sports, culture and politics. However, semantic reference sources, such as Wikidata, DBpedia and EventKG knowledge graphs, provide only an incomplete event series representation. In this paper we target the problem of event series completion in a knowledge graph. We address two tasks: 1) prediction of sub-event relations, and 2) inference of real-world events that happened as a part of event series and are missing in the knowledge graph. To address these problems, our proposed supervised HapPenIng approach leverages structural features of event series. HapPenIng does not require any external knowledge - the characteristics making it unique in the context of event inference. Our experimental evaluation demonstrates that HapPenIng
outperforms the baselines by 44 and 52 percentage points in terms of precision for the sub-event prediction and the inference tasks, correspondingly.
\end{abstract}


\section{Introduction}
\label{sec:introduction}

Event series, such as sports tournaments, music festivals and political elections are sequences of recurring events. 
Prominent examples include the Wimbledon Championships, the Summer Olympic Games, the United States presidential elections and the International Semantic Web Conference. 
The provision of reliable reference sources for event series is of crucial importance for many real-world applications, for example in the context of Digital Humanities and Web Science research \cite{swan2000automatic, GottschalkBRD18, gottschalk2018eventkgtl}, as well as media analytics and digital journalism \cite{mishra2016leveraging, setty2017modeling}.

Popular knowledge graphs (KGs) such as Wikidata \cite{Vrandecic:2012}, DBpedia \cite{dbpedia-swj} and EventKG \cite{Gottschalk:2018, Gottschalk:2019} cover event series only to a limited extent. This is due to multiple reasons:
First, entity-centric knowledge graphs such as Wikidata and DBpedia do not sufficiently cover events and their spatio-temporal relations \cite{Faerber:2015}. 
Second, reference sources for knowledge graphs such as Wikipedia often focus on recent and current events to the detriment of past events \cite{Kaltenbrunner:2012}. 
This leads to the deficiency in supporting event-centric applications that rely on knowledge graphs.

In this work we tackle a novel problem of event series completion in a knowledge graph. 
In particular, we address two tasks: 
1) We predict missing sub-event relations between events existing in a knowledge graph; 
and 2) We infer real-world events that happened within a particular event series but are missing in the knowledge graph. 
We also infer specific properties of such inferred events
such as a label, a time interval and locations, where possible.
Both addressed tasks are interdependent. 
The prediction of sub-event relations leads to a more complete event series structure, facilitating inference of further missing events. 
In turn, event inference can also lead to the discovery of new sub-event relations.

The proposed \approach{} approach exclusively utilizes information obtained from the knowledge graph, without referring to any external sources. 
This characteristic makes \approach{} approach unique with respect to the event inference task. 
In contrast, related approaches that focus on the knowledge graphs population depend on 
external sources (e.g. on news \cite{kuzey2014fresh, yuan2018open}).

\noindent The contributions of this paper include: 
\begin{itemize}
\item A novel supervised method for sub-event relation prediction in event series. 
\item An event inference approach to infer real-world events missing in an event series in the knowledge graph and properties of these events. 
\item A dataset containing new events and relations inferred by \approach{}: 
\begin{itemize}
\item over $5,000$ events and nearly $90,000$ sub-event relations for Wikidata, and
\item over $1,000$ events and more than $6,000$ sub-event relations for DBpedia.
\end{itemize}
\end{itemize}

Our evaluation demonstrates that the proposed \approach{} approach achieves a precision of 
$61\%$ for the sub-event prediction task (outperforming the state-of-the-art embedding-based baseline by $52$ percentage points) and $70\%$ for the event inference task (outperforming a naive baseline by $44$ percentage points). 
Our dataset with new sub-event relations and inferred events is available online\footnote{\url{http://eventkg.l3s.uni-hannover.de/happening}}.

\begin{figure}[t]
\centering
\includegraphics[width=0.85\textwidth]{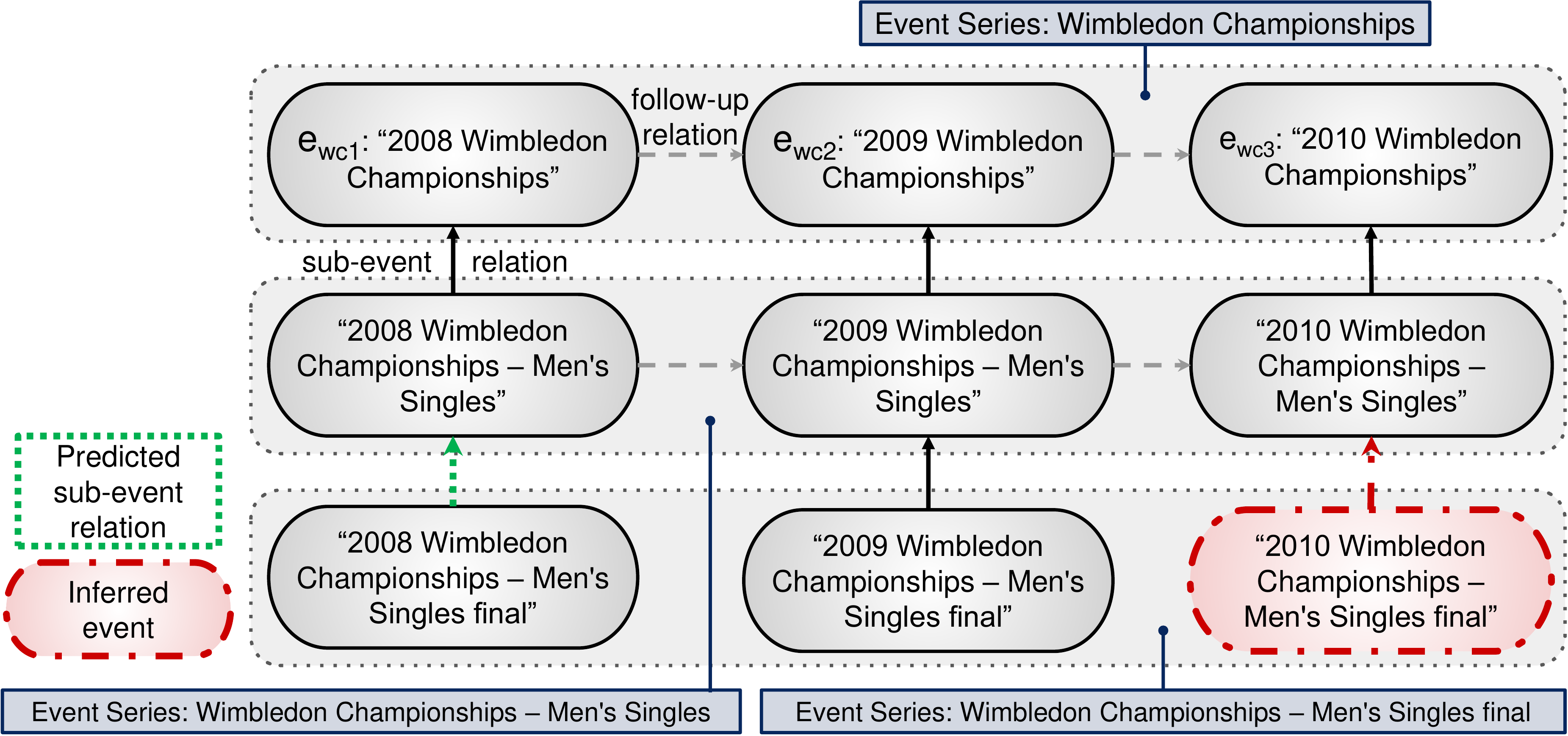}
\caption{A fraction of the \eventGraph{} containing the Wimbledon Championships (WC) events. Nodes represent events. 
Solid arrows represent sub-event relations. Dashed arrows represent follow-up event relations. The three upper events are the $WC$ editions.}
\label{fig:example_wimbledon}
\end{figure}

\subsection{Example: Wimbledon Championships}
\label{sec:example}

The Wimbledon Championships (WC), a famous tennis tournament, are an \textit{event series} that takes place in London annually since 1877. 
Wikidata currently includes $132$ WC editions and $915$ related sub-events, for example Women's and Men's Singles and Wheelchair competitions.
However, according to our analysis, this event series is incomplete. 
In particular, the \approach{} approach proposed in this paper 
was able to generate 
$125$ sub-event relations and $15$ event instances related to this event series
that are currently missing in Wikidata.

Fig. \ref{fig:example_wimbledon} illustrates a small fraction of the \textit{Event Graph} that contains event nodes and their relations as available in Wikidata as of Sep. $18^{th}$, 2018. For each year, Wikidata includes an event edition, such as the \textit{2008 WC}. 
The individual competitions such as the \textit{Men's Singles} are provided as sub-events of the corresponding edition.

In this example we can illustrate two tasks of the event series completion tackled in this paper: (i) Sub-event prediction: The missing sub-event relation between the Men's Singles final of 2008 and the Men's Singles competition in 2008 can be established; and 
(ii) Event inference: The missing event instance labeled \textit{2010 WC --- Men's Singles final} can be inferred as a sub-event of the Men's Singles 2010.

\section{Problem Statement}
\label{sec:problem_statement}

We consider a typical \knowledgeGraph{} that contains nodes representing real-world entities and events. The edges of a \knowledgeGraph{} represent relations between entities and events. More formally:

\definitionskip{}
\begin{definition}
\textbf{\knowledgeGraph{}:} A \knowledgeGraph{} $KG: \langle V, U \rangle$ is a directed multi-graph. 
The nodes in $V$ represent real-world entities and events. The directed edges in $U$ represent relations of the entities and events in $V$.
 \end{definition}
\definitionskip{}

The \eventGraph{} $G$ is a sub-graph of the \knowledgeGraph{}. The nodes of $G$ represent real-world events. The edges represent their relations relevant in the context of event series (sub-event and follow-up relations). More formally:

\definitionskip{}
\begin{definition}
\textbf{\eventGraph{}:} 
Given a \knowledgeGraph{} $KG: \langle V, U \rangle$, an \eventGraph{} $G:$ $\langle$ $E$, $R \cup F$ $\rangle$ is a directed graph. The nodes of the \eventGraph{} $E \subseteq V$ represent real-world events. The edges $R$ represent sub-event relations: $R \subseteq E \times E, R \subseteq U$. The edges $F$ represent follow-up event relations: $F \subseteq E \times E, F \subseteq U$.
\end{definition}
\definitionskip{}

Events in $G$ represent real-world happenings; the key properties of an event in the context of event series include an event identifier, an event label, a happening time interval and relevant locations.  

\definitionskip{}
\begin{definition}
\textbf{Event:} 
Given an \eventGraph{} $G:\langle E, R \cup F \rangle$,
an \textit{event} $e \in E$  is something that happened in the real world. $e$ is represented as a tuple $e = \langle uri, l, t, L\rangle$, where $uri$ is an event identifier, $l$ is an event label, $t=\langle t_s,t_e\rangle$ is the happening time interval with $t_s$, $t_e$ being its start and end time. $L$ is the set of event locations.
\end{definition}
\definitionskip{}

An event can have multiple sub-events. For example, the \textit{WC Men's single final 2009} is a sub-event of \textit{2009 WC}.

\definitionskip{}
\begin{definition}
\textbf{Sub-event:} 
An event $e_s\in E$ is a \textit{sub-event} of the event $e_p\in E$, i.e. ($e_s,e_p) \in R$, if $e_s$ and $e_p$ are topically related and $e_s$ is narrower in scope. 
\label{def:sub-event}
\end{definition}
\definitionskip{}

We refer to $e_p$ as a parent event of $e_s$. Typically, $e_s$ happens in a temporal and a geographical proximity of $e_p$. 

An event can be a part of an event series. 
An example of an event series is the \textit{WC} that has the \textit{2008 WC} as one of its editions. 

\definitionskip{}
\begin{definition}
\textbf{Event series and editions:} 
An event series $s=\langle e_1, e_2, \ldots, \allowbreak e_n \rangle$, $\forall e_i \in s: e_i \in E$, is a sequence of topically related events that occur repeatedly in a similar form. 
The sequence elements are ordered by the event start time and are called \textit{editions}. We refer to the set of event series as $S$.
\label{def:event_series_and_editions}
\end{definition}
\definitionskip{}

The follow-up relations $F$ connect event editions within an event series. For example, the \textit{2009 WC} is the follow-up event of the 
\textit{2008 WC}.

\definitionskip{}
\begin{definition}
\textbf{Follow-up relation:} 
Given an event series $s=\langle e_1, e_2, \ldots, e_n \rangle$, $e_j$ is a \textit{follow-up} event of $e_i$, i.e. ($e_i,e_j) \in F$, if $e_i \in s$ and $e_j \in s$ are the neighbor editions in $s$ and $e_i$ precedes $e_j$.
\end{definition}
\definitionskip{}

The sub-event relations in an \eventGraph{} are often incomplete. In particular, 
we denote the set of real-world sub-event relations not included in the \eventGraph{} as $R^+$. Then the task of sub-event prediction can be defined as follows:

\definitionskip{}
\begin{definition}
\textbf{Sub-event prediction:} 
Given an \eventGraph{} $G: \langle E, R \cup F \rangle$ and events $e_s \in E$, $e_p \in E$, the task of sub-event prediction is to decide if $e_s$ is a sub-event of $e_p$, i.e. to determine if 
$(e_s, e_p) \in R \cup R^+$, where $R^+$ is a set of real-world sub-event relations not included in the \eventGraph{}.
\end{definition}
\definitionskip{}

The set of real-world event representations included in an \eventGraph{} is often incomplete (open world assumption). The context of event series can help to infer real-world events missing in particular editions.  

\definitionskip{}
\begin{definition}
\textbf{Event inference:}
Given and \eventGraph{}  $G: \langle E, R \cup F \rangle$ and an event series $s=\langle e_1, e_2, \ldots, e_n \rangle$, with $e_1, e_2, ..., e_n \in E$, 
the task of event inference is to identify a real-world event $ e_f \in E \cup E^+$that belongs to the series $s$. Here, $ E^+$ is a set of real-world events that are not included in the \eventGraph{}. In particular, $e_f$ is a sub-event of the edition $e_i\in s$, i.e. $(e_f,e_i)\in  R \cup R^+ $.
\end{definition}
\definitionskip{}

\section{Event Series Completion}
\label{sec:approach}

\begin{figure}[t]
\centering
\includegraphics[width=0.75\textwidth]{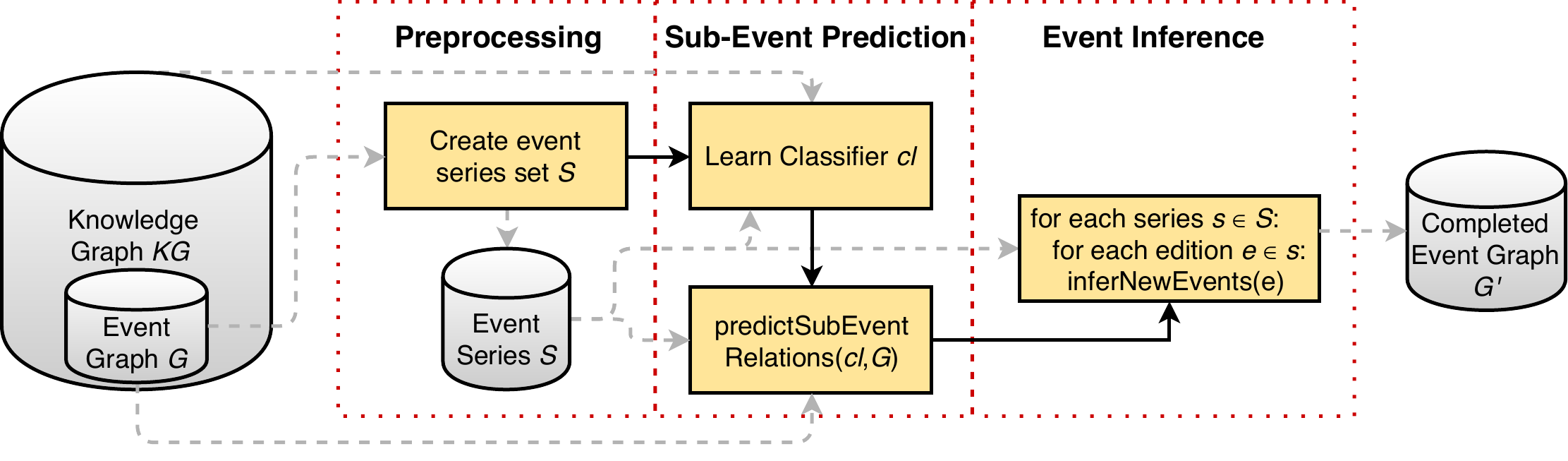}
\caption{The \approach{} pipeline. Solid arrows represent the processing order. Dashed arrows represent the data flow.}
\label{fig:pipeline}
\end{figure}

We address event series completion in two steps: First, we adopt a classification method to predict sub-event relations among event pairs. Second, we develop a graph-based approach to infer events missing in particular editions through event series analysis. 
A pipeline of the overall approach is shown in Fig. \ref{fig:pipeline}.

\subsection{Sub-Event Prediction}
\label{sec:prediction}

We model the problem of sub-event prediction as a classification problem. Given an event pair $(e_s,e_p)$, we aim to predict whether $e_s$ is a sub-event of $e_p$:

\definitionskip{}
\begin{equation}
sub-event(e_s,e_p)=
\begin{cases}
true, & \text{if} ( e_{s}, e_{p}) \in R \cup R^{+}; \\
false, & \text{otherwise.}
\end{cases}
\end{equation}
\definitionskip{}

\subsubsection{Features}
\label{subsubsec:features}
We adopt textual, spatio-temporal and embeddings features.

\textbf{Textual features (\textualFeatures{}):} 
Events connected through a sub-event relation can have similar or overlapping labels whose similarity is measured using textual features. 
Such features are also applied on the \textit{template labels}. Template labels are 
series labels obtained from the original event labels after removal of any digits. 
The textual features we consider include: 

\begin{itemize}
\item Label Containment: $1$, if $e_p.l$ is a sub-string of $e_s.l$, $0$ otherwise.
\item LCS Fraction: The length of the Longest Common Sub-string (LCS) of $e_s.l$ and $e_p.l$, compared to the shorter label: $f_{\textnormal{LCS Fraction}}(e_s,e_p)=\frac{LCS(e_s.l,e_p.l)}{min(|e_s.l|,|e_p.l|)}$.
\item Unigram Similarity: The labels of both events are split into word unigrams. The feature value is the Jaccard similarity between the unigram sets: \\ $f_{\textnormal{Unigram Similarity}}(e_s,e_p)=\frac{\textnormal{unigrams}(e_s.l)\ \cap\ \textnormal{unigrams}(e_p.l)}{\textnormal{unigrams}(e_s.l)\ \cup\ \textnormal{unigrams}(e_p.l)}$.
\item Template Containment, Template LCS Fraction, Template Unigram Similarity: 
These features are computed equivalent to the label features, but are based on the template labels.
\item Label Cosine Similarity: The cosine similarity between event labels based on tf-idf vectors to take frequency and selectivity of terms into account.
\item Parent Event Label Length: 
$f_{\textnormal{Parent Event Label Length}}(e_s,e_p)= |e_p.l|$.
\item Sub-Event Label Length: 
$f_{\textnormal{Sub-Event Label Length}}(e_s,e_p)= |e_s.l|$.

\end{itemize}

\textbf{Spatio-temporal features (\spatioTemporalFeatures{}):} We assume that sub-events happen in the temporal proximity of their parent events.
We consider the temporal proximity through temporal overlap, containment and equality.

\begin{itemize}
\item Time Overlap: $1 \textnormal{ if } e_s.t \cap e_p.t \neq \varnothing$, $0$ otherwise.
\item Time Containment: $1 \textnormal{ if } e_s.t \subseteq e_p.t$, $0$ otherwise.
\item Time Equality: $1 \textnormal{ if } e_s.t = e_p.t$, $0$ otherwise.
\end{itemize}

Sub-events typically happen in the geographical proximity of their parent events. Therefore, we introduce Location Overlap - a spatial feature that assigns a higher score to the event pairs that share locations:

\begin{itemize}
\item Location Overlap: $1 \textnormal{ if } e_s.L \cap e_p.L \neq \emptyset$, $0$ otherwise.
\end{itemize}

\textbf{Embedding features (\embeddingFeatures{}):} 
The link structure of the \knowledgeGraph{} can be expected to provide important insights into possible event relations. First, we can expect that this structure provides useful hints towards predicting sub-event relations, e.g. follow-up events can be expected to have a common parent event. 
Second, events related to different topical domains (e.g. politics vs. sports)
are unlikely to be related through a sub-event relation.
To make use of this intuition, we train an embedding on the \knowledgeGraph{} using any relations connecting two events in $E$. 
For this feature, we pre-train the embeddings following the STransE embedding model \cite{Nguyen:2016} which provides two relation-specific matrices $\boldsymbol{W_1}$ and $\boldsymbol{W_2}$, a relation vector $\boldsymbol{r}$ and entity vectors (here, $\boldsymbol{e_s}$ and $\boldsymbol{e_p}$). Intuitively, given that model, we can compare the embedding of an event with the embedding of the assumed parent event plus the embedding of the sub-event relation ($sE$):

\begin{itemize}
\item Embedding Score: $f_{\textnormal{Embedding}}(e_s,e_p) = 
\lVert \boldsymbol{W}_{r_{\textnormal{sE}},1}\boldsymbol{e}_p + \boldsymbol{r}_{\textnormal{sE}} - \boldsymbol{W}_{r_{\textnormal{sE}},2}\boldsymbol{e}_s \rVert _{\ell_1}$

\end{itemize}

\subsubsection{Training the sub-event classifier}
\label{subsubsec:training}

To train a classifier given the features presented above, a set of labeled event pairs is required. The set of positive examples contains all event pairs with known sub-event relations in the \eventGraph{} $G$. Formally, given the set $E$ of events, this is the set $C_{+} = \{(e_s,e_p) | (e_s,e_p) \in R\}$.

In addition, a set of negative examples, i.e. event pairs without sub-event relation is required. 
When composing event pairs randomly, most of the paired events would be highly different (e.g. having highly dissimilar labels and no spatio-temporal overlap). Consequently, the model would only learn to distinguish the most simple cases. 
To address this problem, we collect a set of negative examples $C_{-}$ that has as many event pairs as $C_{+}$, and consists of four equally-sized subsets with the following condition for each contained event pair ($e_s,e_p$):
\begin{itemize}
\item Both events are from the same event series, but $(e_s,e_p) \notin R$.
Example: (\textit{1997 WC --- Women's Doubles}, \textit{2009 WC --- Men's Singles final}).
\item Both events have the same parent event.
Example: (\textit{2009 WC --- Men's Singles}, \textit{2009 WC --- Women's Singles}).
\item The parent of $e_s$'s parent is the same as $e_p$'s parent.
Example: (\textit{2009 WC --- Men's Singles final}, \textit{2009 WC --- Women's Singles}).
\item $e_s$ is a transitive, but not a direct sub-event of $e_p$.
Example: (\textit{2009 WC --- Men's Singles final}, \textit{2009 WC}). 
\end{itemize}

Note that we only consider direct sub-event relations to be valid positive examples. In particular, we aim to learn to distinguish the directly connected sub-events from transitive relations, as well as to distinguish similar events that belong to different editions.
Due to the inherent incompleteness of the \eventGraph{}, a missing sub-event relation does not necessarily imply that this relation does not hold in the real world. 
However, we expect that false negative examples would occur only rarely in the training set, such that the resulting model will not be substantially affected by such cases. 

Overall, the set of training and test instances $C$ contains all positive sub-event examples $C_{+}$ found in the \eventGraph{}, and an equally sized set of negative examples $C_{-}$ that consists of the four event pair sets described above.

\subsubsection{Predicting sub-event relations using the classifier}
\label{subsubsec:applying_classifier}
The trained classifier is adopted to predict missing sub-event relations within event series. We apply an iterative algorithm, given a classifier $cl$ and the \eventGraph{} $G$. 
As it is not feasible to conduct a pairwise comparison of all events in $G$, we limit the number of events compared with their potential parent event: 
For each potential parent event $e_p$ that is part of an event series, a set of candidate sub-events is selected as the set of events with the largest term overlap with the potential parent event label. For each candidate event, the classifier $cl$ predicts whether this event is a sub-event of $e_p$. To facilitate prediction of sub-event relations in cases where the parent event is not a part of the series initially, the procedure is run iteratively until no new sub-event relations are found.

\subsection{Event Inference}
\label{sec:inference}

The task of event inference is to infer real-world events not initially contained in the \eventGraph{}  (i.e. events in the set $E^+$). 
We infer such missing events and automatically generate their key properties such as label, 
time frame and location, where possible.
The intuition behind event inference is that the \eventGraph{} indicates certain patterns repeated across editions. 
Thus, we approach this task via comparison of different editions of the same event series to recognize such patterns. Consider the WC example in Fig. \ref{fig:example_wimbledon}. Although there is no event instance for the \textit{2010 Men's Singles final}, we can infer such instance from the previous edition \textit{2009 Men's Singles final}.

\subsubsection{Event Series Pre-processing}
\label{sec:pre-processing}

We pre-process the set $S$ of event series to avoid cycles or undesired dependencies within the single series.
Each event series is transformed into a sequence of acyclic rooted trees where each root represents one particular edition of the series. Events or relations violating that structure are removed from the series. If removal is not possible, we exclude such series from $S$.

An important concept of the event inference is the concept of a sub-series: A series $s_p$ has a sub-series $s_s$ if the sub-series contains sub-events of $s_p$. 
For example, the \textit{WC --- Men's Singles final} series is a sub-series of the \textit{WC --- Men's Singles}, because the event \textit{2009 WC --- Men's Singles final} is a sub-event of \textit{2009 WC --- Men's Singles}. 
We determine sub-series relation as:

\definitionskip{}
\begin{definition}
\textbf{Sub-series:} 
An event series $s_s \in S$ is a \textit{sub-series} of $s_p \in S$, if for an event $e_p \in s_p$ there is a sub-event in $s_s$: $\exists (e_s,e_p) \in R: e_p \in s_p \land e_s \in s_s$.
\label{def:sub-series}
\end{definition}
\definitionskip{}

\subsubsection{Inferring New Events}
\label{sec:inferring_new_events}

\setlength{\textfloatsep}{5pt}
\begin{algorithm}[t]
\caption{Event Inference}\label{alg:sub_event_generation}
\begin{algorithmic}[1]
\Procedure{InferSubEvents}{$e$}

\State $M \gets $ getSubSeries($e.series$)\label{alg:sub_event_generation_M}
\ForAll{$e_s \in \{e_s | (e_s, e) \in R\}$} $M = M \setminus e_s.series$
\EndFor\label{alg:sub_event_generation_m2}

\ForAll{$m \in M$}\label{alg:sub_event_generation_constraints}
\If{constraintsNotSatisfied($m, e$)}\label{alg:sub_event_generation_constraints_start} \textbf{continue} \label{alg:sub_event_generation_constraints_end}
\EndIf
\State $newEvent \gets $ inferEvent($e, m$)
\If{$oldEvent \gets $ findEvent($E$, $newEvent.l$) $\neq \varnothing$}
\State $R = R \cup (e, oldEvent)$ \label{alg:sub_event_exists}
\Else
\State $E = E \cup newEvent$; $R = R \cup (e, newEvent)$
\EndIf
\EndFor\label{alg:sub_event_generation_endfor}

\ForAll{$e_s \in \{e_s | (e_s, e_p) \in R\}$} inferSubEvents($e_s$)
\EndFor
\EndProcedure
\end{algorithmic}
\end{algorithm}

The intuition behind event inference is to identify similar patterns in the different editions of an event series. According to Definition \ref{def:event_series_and_editions}, the editions of an event series occur repeatedly in a similar form. 
This way, events repeated in most of the editions of the series, but missing in a particular edition can be inferred. 
To do so, we process all editions in the \eventGraph{} and inspect whether its neighbored editions have a sub-event not covered in the particular edition.

Algorithm \ref{alg:sub_event_generation} illustrates our event inference approach. As shown in our pipeline (Fig. \ref{fig:pipeline}), this algorithm is invoked for each edition $e$ of the event series in $S$. 
First, a set $M$ is constructed that contains all sub-series of the current edition's series, i.e. 
$e.series$ (line \ref{alg:sub_event_generation_M}). 
Then, the algorithm removes all series from $M$ for which the current edition contains events already
(line \ref{alg:sub_event_generation_m2}). That way, $M$ is reduced to a set of event series not covered by the sub-events of the current edition $e$.

For each remaining sub-series $m \in M$, a new event is inferred that is a sub-event of the current edition $e$ and a part of $m$. Within the respective method \texttt{inferEvent($e,M$)}, a new label, time span and set of locations is generated as described later.
The algorithm is invoked recursively with all known (also newly identified) sub-events.
To increase precision, a sub-series $m$ is only retained in $M$ if a set of constraints is satisfied (line \ref{alg:sub_event_generation_constraints_start}). These constraints are described later in this section.

The event inference algorithm can infer an event for which an equivalent event already exists in the \eventGraph{}. 
To avoid the generation of such duplicate events, we check if an event with the same label as the newly inferred event exists in the \eventGraph{}. In this case, the algorithm adds a new sub-event relation across the existing events to the \textit{Event Graph} and discards the inferred event (line \ref{alg:sub_event_exists}).

\begin{figure}[t]
\centering
\begin{subfloat}[Step 1: The event inference algorithm is invoked with the \textit{2010 WC} event $e$. For the sub-series $m$ of $e.series$, \textit{2010 WC --- Men's Singles} is already a sub-event of $e$. No new event is inferred.]{
  \includegraphics[width=.46\linewidth]{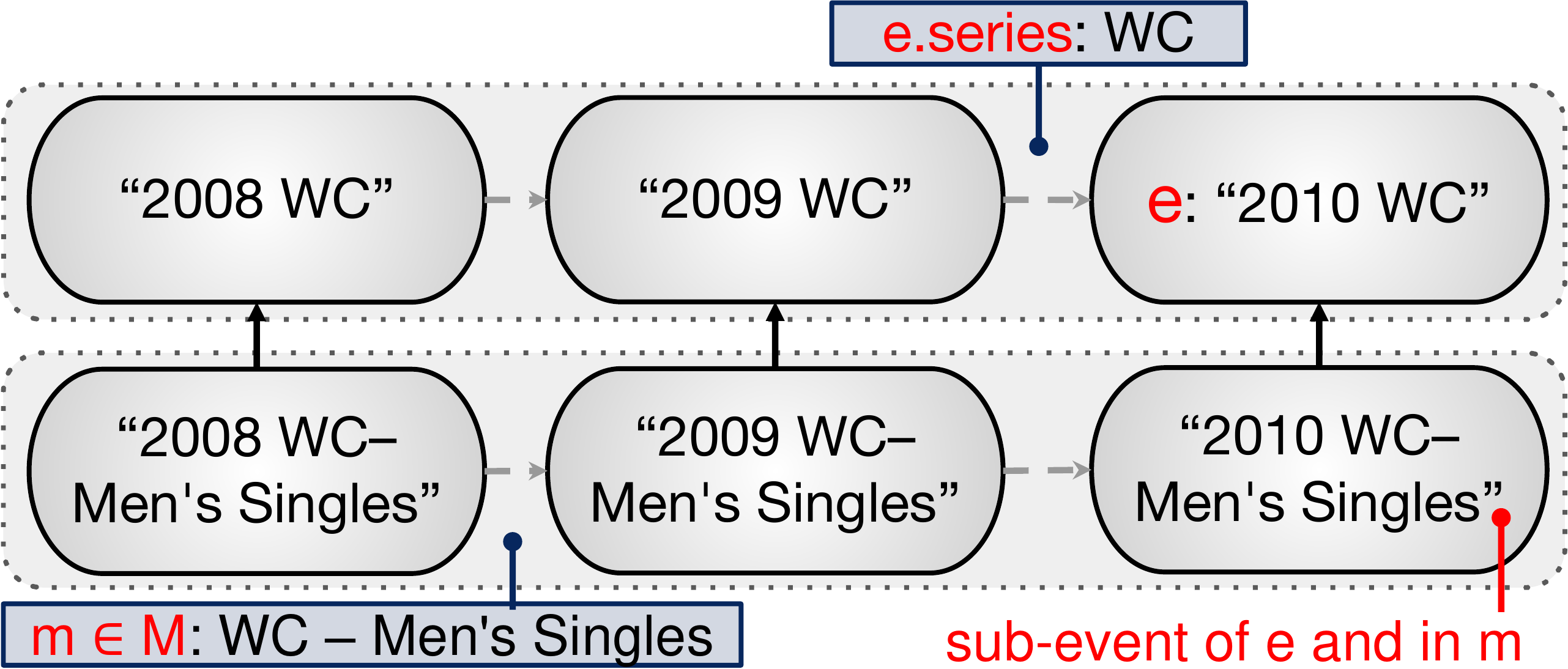}  \label{fig:example_inference_01}
  }
\end{subfloat}%
\quad
\begin{subfloat}[Step 2: The algorithm is now invoked with the \textit{WC 2010 --- Men's Singles} event $e$. For the sub-series $m$ of $e.series$, there is no sub-event of $e$. A new event is inferred.]{
  \includegraphics[width=.46\linewidth]{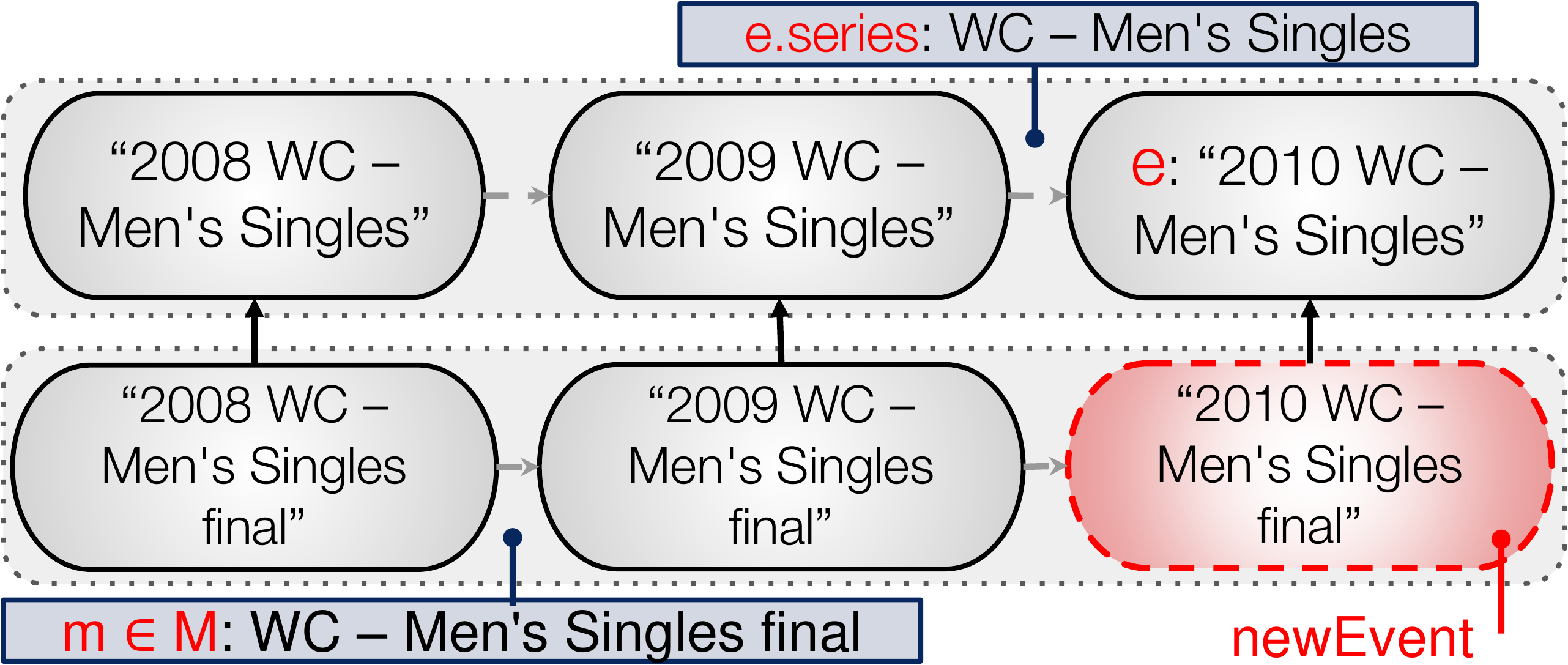}
  \label{fig:example_inference_02}
  }
\end{subfloat}
\caption{Event inference example for the Wimbledon Championships.}
\label{fig:example_inference}
\end{figure}

\subsubsection{Wimbledon Championships Example:} Consider the example in Fig. \ref{fig:example_wimbledon}, with the goal to infer new events within the edition $e_{wc_3}$: \textit{2010 WC}.
Fig. \ref{fig:example_inference_01} depicts the first step when invoking the algorithm \texttt{Infer\-Sub\-Events($e_{wc_3}$)} (without constraints). The edition becomes the input event $e$ and its series $e.series$ is \textit{WC}. The event series \textit{WC --- Men's Singles} ($m$) is identified as one of its immediate sub-series in $M$. However, as there is already an event \textit{2010 WC --- Men's Singles} that is a sub-event of $e$ and part of that sub-series $m$, it is removed from $M$. Therefore, $M$ is empty and no new events are inferred at this point.

Subsequently, Algorithm \ref{alg:sub_event_generation} is executed with the sub-event \textit{2010 WC --- Men's Singles} as input edition $e$, as shown in Fig. \ref{fig:example_inference_02}. Here, the sub-series is \textit{WC --- Men's Singles final} which is inserted in $M$. Consequently, a new event is created that is a sub-event of $e$ and part of the event series \textit{WC --- Men's Singles final}. 

\setlength{\textfloatsep}{5pt}
\begin{algorithm}[t]
\caption{Label Generation}\label{alg:label_generation}
\begin{algorithmic}[1]
\Procedure{GenerateLabel}{$e$, $m$}
\State $mostSimilarEvents \gets $ getSimilarEvents($e$, $e.series$) \label{alg:label_generation:step1}
\State sortEventsByEditionCloseness($e$, $mostSimilarEvents$)\label{alg:label_generation:step2_start}
\State $c \gets mostSimilarEvents[0] $
\State $c' \gets c'$, s.t. $(c',c) \in R \land c' \in m $\label{alg:label_generation:step2_end}
\State $l \gets ""$; $r \gets c'.l; \delta_{prev} \gets \emptyset$ \label{alg:label_generation:step3_start}
\ForAll{$\delta \in $ getEdits($c.l$, $e.l$)}
\If {$\delta.op = \textnormal{DELETE}$} $\delta_{prev} \gets \delta$
\ElsIf {$\delta.op = \textnormal{INSERT} \land \delta_{prev}.op = \textnormal{DELETE}$}
\State{$l \gets l + r[:r.indexOf(\delta_{prev}.text)] + \delta.text$}
\State{$r \gets l + r[r.indexOf(\delta_{prev}.text) +len(\delta_{prev}.text):]$}
\ElsIf {\textbf{not} $(\delta.op = \textnormal{EQUAL} \land \delta_{prev} = \emptyset)$ } \textbf{return} $\emptyset$
\EndIf
\EndFor \label{alg:label_generation:step3_end}
\Return $l + r$
\EndProcedure
\end{algorithmic}
\end{algorithm}

\begin{table}[t]
\centering
\footnotesize
\begin{tabu}{r|[1.5pt]l|l|l|c|l}
\multicolumn{1}{c|[1.5pt]}{\textbf{step}} & \multicolumn{1}{|c|}{\textbf{$\delta$.op}} & \multicolumn{1}{|c|}{\textbf{$\delta_{prev}$.op}} &\multicolumn{1}{|c|}{\textbf{$\delta$.text}} & \multicolumn{1}{|c}{\textbf{l}} & \multicolumn{1}{|c}{\textbf{r}} \\ \tabucline[1.5pt]{-}
init & & & & & 2009 WC - Men's Singles final \\ \hline
1 & DELETE & & 2009 & & 2009 WC - Men's Singles final \\ \hline
2 & INSERT & DELETE & 2010 & 2010 & WC - Men's Singles final \\ \hline
3 & EQUAL & & WC - Men's Singles & 2010 & WC - Men's Singles final
\end{tabu}
\caption{Generating the label \textit{2010 WC - Men's Singles}. The edit operations $\delta$ are the result of Myers' algorithm to detect the edit operations between \textit{2009 WC - Men's Singles} and \textit{2010 WC - Men's Singles}. The final label is the concatenation of $l$ and $r$.}
\label{tab:example_label_generation}
\end{table}

\subsubsection{Label Generation}
\label{subsubsec:label_generation}

Each newly generated event requires a label.
This label is generated by exploiting the labels within its event series, as shown in Algorithm \ref{alg:label_generation}.
The input is its future parent event $e$ and its event series $m$. 
First, the events in the parent series $e.series$ whose labels are most similar to the label of $e$ are collected (line \ref{alg:label_generation:step1}).
Then, within this set of events, the one from the closest event edition and its sub-event in $m$ is selected (lines \ref{alg:label_generation:step2_start} - \ref{alg:label_generation:step2_end}).
Finally, the label of that event is transformed into the new label by applying the same edit operations $\delta$ (i.e. equality, delete or insert) as if we transformed the parent event labels (lines \ref{alg:label_generation:step3_start} - \ref{alg:label_generation:step3_end}). 
To identify the edits, we adopt the difference algorithm by Myers \cite{myers1986ano}.

\textit{Example:} Consider the newly added event in Fig. \ref{fig:example_inference_02}. As an input to the algorithm, there is $e$ which is the event \textit{2010 WC --- Men's Singles} and the series $m$ consisting of the Men's Singles finals of 2008 and 2009. First, the event \textit{2009 WC --- Men's Singles} within $e.series$ is identified as the most similar event $c$. $c'$ is the sub-event of $c$ that is also in $m$: \textit{2009 WC --- Men's Singles final}. Given $c'.l$ and the edit operations $\delta$ between the labels of $e$ and $c$, Table \ref{tab:example_label_generation} shows how they are used to generate the correct label \textit{2010 WC --- Men's Singles final}.

\subsubsection{Location and Time Generation}
\label{sec:infer_loc_and_time}

Each event can be assigned a happening time and a set of locations. In both cases, we use a rule-based approach. 

\textit{Locations:} Some events such as the Olympic Games change their location with every edition. Currently we reconstruct event locations only if they remain unchanged across editions: 
If there is a location assigned to every event $s \in m$, this location is also assigned to $e$. 
In future work we intend to utilize sub-location relations, that facilitate the generation of correct locations at a lower level of geographical granularity.

\textit{Happening Times:} 
Three rules are applied in the following order until a happening time is identified: a) If the happening time of each event $s \in m$ equals its parent event's happening time, also $e$ adopts its happening time directly from its parent event; b) If the happening time of each event $s \in m$ is modelled as a whole year, the happening time of $e$ is also modelled as the same year as any of its (transitive) parent events; c) If the event label contains a year expression, that part is transformed into its happening time.

\subsubsection{Constraints}
\label{subsubsec:constraints}

We propose several configurations of constraints to decide whether an event should be created:

\begin{itemize}
    \item Baseline (\modSimple{}): No constraints.
    \item Time Evolution (\modEvo{}): The constraints are only satisfied if there was at least one event in the series that happened before $e$. For example, the \textit{Wimbledon Women's Doubles} were held for the first time in 1913, so it would be wrong to generate an event for the \textit{Women's Doubles} series in 1912 and before.
    \item Interval (\modInterval{}): The constraints are only satisfied if there was at least one event in the series that happened before and at least one event in the series that happened after $e$. Under this constraint, events that re-occurred only until a specific edition are not generated for each edition. An example is the tug of war which was part of only six Olympic Summer Games.
    \item Window (\modWindow{}): Given a start and an end thresholds $a$ and $b$, this constraint is satisfied if there is at least one event within the last $a$ editions of the series that happened before $e$ and at least one event in the following $b$ editions that happened after $e$. For example, Tennis competitions in the Olympic Summer Games were held between 1896 and 1924, and then only since 1984. The Window constraint helps to identify such gaps.
    \item Coverage (\modCoverage{}): Event series are only valid if they are part of a sufficient fraction of the editions: $\nicefrac{|m|}{|S|} \geq \alpha$, given a threshold $\alpha$.
    \item Coverage Window (\modCovWin{}): A combination of \modWindow{} and \modCoverage{}: The coverage is only computed after restricting both event series to the dynamic time window.
    \item Evolution Coverage Window (\modEvoCovWin{}): A combination of \modEvo{}, \modWindow{} and \modCoverage{}: The coverage is only computed after restricting both event series to the dynamic time window, and if at least one event in the series happened before $e$.
\end{itemize}

\section{Evaluation}
\label{sec:evaluation}

The goal of the evaluation is to assess the performance of the \approach{} approach with respect to the sub-event prediction and event inference tasks.

\subsection{Data Collection and Event Graph Construction}
\label{sec:data_collection_preprocessing}

We run our experiments on \eventGraphs{} extracted from two sources: (i) Wikidata \cite{Vrandecic:2012} as of October 25, 2018 (\wikidataEventGraph{}), and (ii) DBpedia \cite{dbpedia-swj} from the October 2016 dump (\dbpediaEventGraph{}). Both datasets are enriched with additional information regarding events obtained from the EventKG knowledge graph \cite{Gottschalk:2018}. Compared to other knowledge graphs, EventKG contains more detailed information regarding spatio-temporal characteristics of events. More concretely, events in the \eventGraph{} are enriched with location and time information using the properties \textit{sem:hasPlace}, \textit{sem:hasBeginTimeStamp} and \textit{sem:hasEndTimeStamp} of EventKG.

One \eventGraph{} containing events, sub-event relations and follow-up relations, as well as a set $S$ of event series is constructed for each dataset.
For the \wikidataEventGraph{}, we collect as events all data items that are (transitive) instances of the ``event" class\footnote{\url{https://www.wikidata.org/wiki/Q1656682}}. 
Event series are extracted using the ``instance of''\footnote{https://www.wikidata.org/wiki/Property:P31} and the ``series''\footnote{https://www.wikidata.org/wiki/Property:P179} properties in Wikidata. For the \dbpediaEventGraph{}, we extract events using the ``dbo:Event'' class and series assignments using the provided Wikipedia categories. In both cases, we apply two heuristics to ensure that only event series compatible with Definition \ref{def:event_series_and_editions} are extracted: (i) We only consider series with mostly homogeneous editions. To this end, we make use of the Gini index \cite{raileanu2004theoretical}, a standard measure for measuring impurity. In our context it is used to assess the diversity of the template labels of editions in an event series. We reject the (rare) cases of event series with high Gini impurity, where the edition labels do not follow any common pattern.\footnote{For example, the event series ``TED talk'', whose set of edition template labels (e.g. ``Avi Reichental: What’s next in 3D printing'' and ``Amanda Palmer: The art of asking'') has a high Gini impurity, is not included in the set of event series.} An event is kept in $S$, if the set of template labels of its editions shows a Gini impurity less than $0.9$. Besides, we ignore editions whose removal decreases that impurity. (ii) We ignore events typed as military conflicts and natural disasters, because such events typically do not follow any regularity.
If we can find connected sub-graphs of events in the \eventGraph{} through sub-event and follow-up relations, but the data item representing that series is missing in the dataset, we add a new unlabeled event series to $S$. To train the embeddings, we collect all relations connected to events.

The extraction process results in a \wikidataEventGraph{} $G$ containing $|E|=352,235$ events (\dbpediaEventGraph{}: $92,523$) and $|S|=9,007$ event series (\dbpediaEventGraph{}: $1,871$). As input to train the embeddings, there are $279,004,908$ relations in Wikidata and $18,328,678$ relations in DBpedia. Both \eventGraphs{}, as well as embeddings, annotated samples and other evaluation datasets described in the remainder of this section are available online.\footnote{\url{http://eventkg.l3s.uni-hannover.de/happening}}

\subsection{Sub-Event Prediction}
\label{sec:sub_event_prediction}

\subsubsection{Training and Test Set Generation}
\label{sec:training_and_test_set_generation}

Before running the experiments, a set of positive and negative sub-event relations is created from the \eventGraphs{} as described in Section \ref{subsubsec:training}. 
In total, this collection of relations consists of $55,217$ event pairs within $S$ that were extracted as correct sub-event pairs from Wikidata (DBpedia: $16,763$) and the same number of negative event pairs.\footnote{Existing benchmark datasets do not contain a sufficient amount of sub-event relations. For example, FB15K \cite{Bordes:2013} only contains $224$ triples containing one of the Freebase predicates \textit{/time/event/includes\_event}, \textit{/time/event/included\_in\_event} or \textit{/time/event/instance\_of\_recurring\_event}.} This collection is split into ten folds to allow 10-fold cross-validation. 
We learn the STransE embeddings as described in Section \ref{subsubsec:features} for each fold, with its parameters set as follows: SGD learning rate $\lambda = 0.0001$, the margin hyper-parameter $\gamma = 1$, vector size $k = 100$ and $1,000$ epochs. While learning the embeddings on the folds, we exclude the sub-event relations from the respective test set. 

\subsubsection{Baseline}
\label{sec_baseline}

As a baseline for sub-event prediction, we utilize an embedding-based link prediction model based on the STransE embeddings \cite{Nguyen:2016}. Given an input event, this model retrieves a ranked list of candidate sub-events with the corresponding scores. We use these scores to build a logistic regression classifier. STransE is a state-of-the-art approach that had been shown to outperform previous embedding models for the link prediction task on the FB15K benchmark \cite{Bordes:2013}.

\subsubsection{Classifier Evaluation}
\label{sec:classifier_evaluation}

\begin{table}[t]
\centering
\begin{tabu}{ll|[1.5pt]r|r|r|r|r|[1.5pt]r}
& & \multicolumn{5}{c|[1.5pt]}{\textbf{Wikidata}} & \multicolumn{1}{c}{\textbf{DBpedia}}  \\ \cline{3-8}
\multicolumn{2}{c|[1.5pt]}{\textbf{Method}} & \multicolumn{1}{c|}{\textbf{TP}} & \multicolumn{1}{c|}{\textbf{TN}} & \multicolumn{1}{c|}{\textbf{FP}} & \multicolumn{1}{c|}{\textbf{FN}} & \multicolumn{1}{c|[1.5pt]}{\textbf{Accuracy}} & \multicolumn{1}{c}{\textbf{Accuracy}} \\ \tabucline[1.5pt]{-}
Baseline & \stranse{} & 46,479 & 43,143 & 6,949 & 13,859 & 0.81 & 0.50 \\ \tabucline[1.5pt]{-}
\multirow{3}{*}{\begin{tabular}[c]{@{}l@{}}\approach{} \\ configurations\ \ \ \end{tabular}} & \logReg{} & 54,345 & 46,605 & 3,487 & 5,993 & 0.91 & 0.87 \\ \cline{2-8}
& \svm{} & 55,958 & 48,825 & 1,267 & 4,380 & 0.95 & 0.92 \\ \cline{2-8}
& \randomForest{} & \textbf{58,649} & \textbf{49,497} & \textbf{595} & \textbf{1,689} & \textbf{0.98} & \textbf{0.97} 
\end{tabu}
\caption{10-fold cross-validation of the sub-event prediction using different classifiers and all the introduced features. \stranse{} is the baseline we compare to.}
\label{tab:evaluation_classification}
\end{table}

Table \ref{tab:evaluation_classification} shows the results of the 10-fold cross-validation for the sub-event prediction task, with three different classifiers: \logReg{} (Logistic Regression), \randomForest{} (Random Forest) and \svm{} (Support Vector Machine with linear kernel and normalization) in terms of classification accuracy ($\frac{TP + TN}{TP + TN +FP + FN}$, where $TP$ are true positives, $TN$ true negatives, $FP$ false positives and $FN$ false negatives).
Among our classifiers, the \randomForest{} classifier performs best, with an accuracy of nearly $0.98$ in the case of the \wikidataEventGraph{} and $0.97$ for the \dbpediaEventGraph{}. The results show a clear improvement over the \stranse{} baseline, outperforming the baseline by more than $16$ percentage points in case of the \randomForest{} classifier for Wikidata.
For DBpedia, the \stranse{} baseline is outperformed by a larger margin using our proposed features. This can be explained by the insufficient number of relations for training the embeddings in DBpedia.

Table \ref{tab:feature_groups_classification} shows the performance of the \randomForest{} classifier under cross-validation with different feature groups. The combination of all features leads to the best performance in terms of accuracy. Although the use of textual features already leads to a high accuracy ($0.97$), embedding features and spatio-temporal features help to further increase accuracy in the case of Wikidata ($0.98$). Again, while DBpedia does profit from the spatio-temporal features, there is no improvement when using embeddings, due to the insufficient data size.

\begin{table}[t]
\centering
\setlength\tabcolsep{4pt} 
\begin{tabu}{l|[1.5pt]r|[1.5pt]r}
\multicolumn{1}{c|[1.5pt]}{\textbf{Feature Group}} &  \begin{tabular}[c]{@{}l@{}}\textbf{Wikidata} \\ \textbf{Accuracy}\end{tabular} & \begin{tabular}[c]{@{}l@{}}\textbf{DBpedia} \\ \textbf{Accuracy}\end{tabular} \\ \tabucline[1.5pt]{-}

All features: \textualFeatures{}, \spatioTemporalFeatures{}, \embeddingFeatures{} & \textbf{0.98} & 0.97  \\ \hline

No spatio-temp. features: \textualFeatures{}, \embeddingFeatures{} & 0.97 & 0.96  \\ \hline

No textual features: \spatioTemporalFeatures{}, \embeddingFeatures{} & 0.82 & 0.73  \\ \hline
 
No embedding: \textualFeatures{}, \spatioTemporalFeatures{} & 0.98 & \textbf{0.97} \\
\end{tabu}
\caption{10-fold cross-validation of the sub-event prediction using the \randomForest{} classifier for Wikidata and DBpedia with different feature sets.}
\label{tab:feature_groups_classification}
\end{table}

\subsubsection{Wikidata Statistics and Examples}
\label{sec:wikidata_statistics_prediction}

While the classifiers demonstrate very accurate results on the test sets, the performance on predicting sub-event relations not yet contained in $G$ requires a separate evaluation. As explained in Section \ref{subsubsec:applying_classifier}, a large number of predictions is needed that could potentially also lead to a large number of false positives, even given a highly accurate classifier. The actual label distribution is skewed towards unrelated events and we are now only classifying event pairs not yet contained in $R$.
In fact, running the sub-event prediction algorithm using the best-performing \randomForest{} classifier with all features leads to the prediction of $85,805$ new sub-event relations not yet contained in Wikidata and $5,651$ new sub-event relations in DBpedia.

To assess the quality of the predicted sub-event relations that are not initially contained in $R$, we extracted a random sample of $100$ sub-event relations consisting of an event and its predicted sub-event and manually annotated each pair as correct or incorrect sub-event relation. According to this manual annotation, $61\%$ of the sub-event relations predicted with our \approach{} approach that are not yet contained in the \eventGraph{} correctly represent real-world sub-event relations in Wikidata (DBpedia: $42\%$). In comparison, the \stranse{} baseline predicted only $46,807$ new sub-event relations, and only $9\%$ of them are correct based on a manual annotation of a random 100 relations sample (DBpedia: $2\%$).
The difference in performance on the test set and on the predicted sub-event relations not contained in $R$ can be explained by the large class disbalance in the set of relations collected in the sub-event prediction procedure, such that the majority of the candidate relations are negative examples.

\subsection{Event Inference Performance}
\label{sec:event_inference_performance}

We evaluate the event inference performance in two steps: First, we conduct an automated evaluation of recall by reconstruction of corrupted event series. Second, we assess precision by annotating random samples of new events.

\begin{table}[t]
\centering
\small
\begin{tabu}{lc|[1.5pt]r|r|r|[1.5pt]r|r|r}
& & \multicolumn{3}{c|[1.5pt]}{\textbf{Wikidata}} & \multicolumn{3}{c}{\textbf{DBpedia}} \\ \cline{3-8} 
& & \multicolumn{6}{c}{\textbf{Corruption Factor}} \\ \cline{3-8} 
\multicolumn{2}{c|[1.5pt]}{\textbf{Constraints}} & \textbf{5\%} & \textbf{10\%} & \textbf{15\%} & \textbf{5\%} & \textbf{10\%} & \textbf{15\%} \\ \tabucline[1.5pt]{-}
Baseline & \multicolumn{1}{c|[1.5pt]}{\textbf{\modSimple{}}} & 
61.81 & 63.13 & 61.83 & 
39.58 & 38.40 & 38.17 \\ \tabucline[1.5pt]{-}
\multirow{6}{*}{\begin{tabular}[c]{@{}l@{}}\approach{} \\ configurations\ \ \ \end{tabular}} & \multicolumn{1}{c|[1.5pt]}{\textbf{\modEvo}} & 
53,63 & 54.70 & 53.12 &
31.04 & 31.32 & 30.12  \\ \cline{2-8}
& \multicolumn{1}{c|[1.5pt]}{\textbf{\modInterval}} & 
46.68 & 47.89 & 46.39 &
24.58 & 24.04 & 23.46  \\ \cline{2-8}
& \multicolumn{1}{c|[1.5pt]}{\textbf{\modWindow}} & 
46.06 & 47.45 & 45.94 &
22.71 & 22.27 & 21.93  \\ \cline{2-8}
& \multicolumn{1}{c|[1.5pt]}{\textbf{\modCoverage}} & 
45.49 & 45.65 & 43.64 &
11.46 & 11.03 & 9.30 \\ \cline{2-8}
& \multicolumn{1}{c|[1.5pt]}{\textbf{\modCovWin}} & 
53.36 & 53.93 & 51.32 &
23.96 & 21.96 & 19.43 \\ \cline{2-8}
& \multicolumn{1}{c|[1.5pt]}{\textbf{\modEvoCovWin}} & 
48.89 & 49.17 & 47.03 &
21.67 & 20.71 & 18.18 
\end{tabu}
\caption{Complementing corrupted event series. For each corruption factor (i.e. \% of removed events), we report the percentage of events that could be reconstructed.}
\label{tab:intrusion} 
\end{table}

\subsubsection{Complementing Corrupted Event Series (Recall)}
\label{sec:complementing_corrupted_event_series}

To evaluate the recall of the event series completion, we remove events from the event series and investigate to which extent our \eventGraph{} completion constraints are able to reconstruct them (we consider the naive unconstrained approach \modSimple{} as our baseline).
To this end, we randomly remove leaf nodes (events without sub-events) from the whole set of event series $S$ until a specific percentage (determined by the \textit{corruption factor}) of leaf nodes is removed. For the \wikidataEventGraph{}, there are $45,203$ such leaf events in total before corruption, for DBpedia $9,600$.
Table \ref{tab:intrusion} shows the results for three corruption factors ($5\%$, $10\%$ and $15\%$) and the constraints introduced in Section \ref{subsubsec:constraints} (we set the parameters to $a=b=5$ and $\alpha = 0.5$). 
As expected, the unconstrained naive approach \modSimple{} results in the highest percentage of correctly reconstructed events: More than $60\%$ of the Wikidata and nearly $40\%$ of the DBpedia events can be recovered including their correct labels. 
If applying constraints, less events are reconstructed. 
In particular, the \modWindow{} constraint results in the lowest recall, as it demands to cover the event before and after the series edition within $5$ editions. 

Overall, we observe that \approach{} is able to reconstruct more than $60\%$ of missing events from a knowledge graph and correctly infer event labels.

\subsubsection{Manual Assessment (Precision)}
\label{sec:manual_assessement}

To access precision, we created random samples of $100$ newly inferred events for each of the constraints proposed in Section \ref{subsubsec:constraints} and both \eventGraphs{}, and manually annotated their correctness. 
Table \ref{tab:event_inference_accuracy} provides an overview of the results. While the naive unconstrained approach results in a precision of less than $0.30$ for both \eventGraphs{}, the inclusion of constraints leads to clear improvement, with a precision of up to $0.70$ for the \modEvoCovWin{} constraint for Wikidata and $0.71$ for the \modWindow{} constraint for DBpedia. 
Table \ref{tab:event_inference_accuracy} also reports the number of additional sub-event relations created during the event inference procedure when checking for duplicate events.

\begin{table}[t]
\centering
\small
\begin{tabu}{lc|[1.5pt]r|r|r|[1.5pt]r|r|r}
 & & \multicolumn{3}{c|[1.5pt]}{\textbf{Wikidata}} & \multicolumn{3}{c}{\textbf{DBpedia}} \\ \cline{3-8}
& & \multicolumn{2}{c|}{\textbf{Inferred Events}} & \multirow{2}{*}{\begin{tabular}[c]{@{}l@{}}\textbf{Rela-}\\ \textbf{tions}\end{tabular}} & \multicolumn{2}{c|}{\textbf{Inferred Events}} & \multirow{2}{*}{\begin{tabular}[c]{@{}l@{}}\textbf{Rela-}\\ \textbf{tions}\end{tabular}} \\ \cline{3-4} \cline{6-7}
 \multicolumn{2}{c|[1.5pt]}{\textbf{Constraints}} & \multicolumn{1}{c|}{\textbf{Number}} & \multicolumn{1}{c|}{\textbf{P}} &  & \multicolumn{1}{c|}{\textbf{Number}} & \multicolumn{1}{c|}{\textbf{P}} & \\ \tabucline[1.5pt]{-}
Baseline & \multicolumn{1}{l|[1.5pt]}{\textbf{\modSimple}} & 
\textbf{114,077} & 0.26 & \textbf{16,877} & \textbf{31,410} & 0.24 & \textbf{3,420} \\ \tabucline[1.5pt]{-}
\multirow{6}{*}{\begin{tabular}[c]{@{}l@{}}\approach{} \\ configurations\ \ \ \end{tabular}} & \multicolumn{1}{l|[1.5pt]}{\textbf{\modEvo}} & 
28,846 & 0.47 & 10,045 & 
11,295 & 0.35 & 1,170 \\ \cline{2-8}
& \multicolumn{1}{l|[1.5pt]}{\textbf{\modInterval}} & 
5,256 & 0.57 & 5,376 & 
2,115 & 0.67 & 3,419 \\ \cline{2-8}
& \multicolumn{1}{l|[1.5pt]}{\textbf{\modWindow}} & 
3,363 & 0.56 & 4,547 & 
936 & \textbf{0.71} & 783 \\ \cline{2-8}
& \multicolumn{1}{l|[1.5pt]}{\textbf{\modCoverage}} & 
7,297 & 0.54 & 2,712 & 
1,313 & 0.45 & 417 \\ \cline{2-8}
& \multicolumn{1}{l|[1.5pt]}{\textbf{\modCovWin}} & 
7,965 &  0.59 & 4,442 & 
1,965 & 0.61 & 718 \\ \cline{2-8}
& \multicolumn{1}{l|[1.5pt]}{\textbf{\modEvoCovWin}} & 
5,010 & \textbf{0.70} & 3,687 & 
1,364 & 0.70 & 655
\end{tabu}
\caption{Manual evaluation of the correctness of inferred events. For the baseline, each \approach{} constraint and \eventGraph{}, $100$ inferred events were randomly sampled and judged as correct or not. The number of additional sub-event relations found during the event inference process is reported as well (P: Precision).}
\label{tab:event_inference_accuracy} 
\end{table}

\subsubsection{Discussion and Additional Statistics}
\label{sec:wikidata_statistics_inference}

The manual assessment shows that \approach{} with the \modEvoCovWin{} constraints is able to infer $5,010$ new events with a precision of $70\%$ in Wikidata and $1,364$ new DBpedia events with similar precision. Events are inferred wrongly in cases where sub-events are happening in an irregular manner. This includes e.g. the wrongly inferred event ``1985 Australian Open -- Mixed Doubles'' that was extracted although there were no Mixed Doubles in that event series between 1970 and 1985 or competitions like the men's single scull in the World Rowing Championships that used to follow a highly unsteady schedule. In future, external knowledge can be used to verify the inferred events.
Differences between the Wikidata and the DBpedia results can be explained by the less complete event type assignments and the lack of a proper sub-event relation in DBpedia, where we use category assignments instead.

As the \modEvoCovWin{} constraint is most precise for the \wikidataEventGraph{}, we provide more insights for this constraint and \eventGraph{} in the following:

\begin{itemize}[topsep=0pt]
    \item Impact of the sub-event prediction on the event inference: If the sub-event prediction step is skipped, only $3,558$ new events are inferred, compared to $5,010$ events otherwise.
    \item Additional relations: $3,687$ new sub-event relations were created during the event inference step in addition to the $85,805$ sub-event relations from the sub-event prediction step (in total: $89,492$ new sub-event relations).
    \item Happening times: $99.36\%$ of the inferred events are assigned a happening time. $0.38\%$ of them were inferred by the first, $81.52\%$ by the second and $18.10\%$ by the third rule from Section  \ref{sec:infer_loc_and_time}.
    \item Locations: Only $79$ of the $5,010$ inferred events were assigned a location under the strict conditions proposed in Section \ref{sec:infer_loc_and_time}.
\end{itemize}

Overall, the two steps sub-event prediction and event inference enable \approach{} to generate ten thousands of new sub-event relations and events. These relations and new instances can be given as a suggestion to be inserted in the respective dataset using human confirmation with external tools, such as the Primary Sources Tool for Wikidata \cite{pellissier2016freebase}.

\section{Related Work}
\label{sec:related_work}

\textbf{Knowledge Graph Completeness}. Completeness is an important dataset quality dimension \cite{EllefiBBDDST18}. Due to the \textit{open-world assumption} knowledge graphs are notoriously incomplete. 
The facts not present in the knowledge graph are unknown, and may or may not be true \cite{razniewski2016but, tanon2017completeness}.
There has been research on several exemplary aspects of knowledge graph completeness, for example on the incompleteness of Wikidata \cite{Balaraman:2018, Ahmeti:2017} and the relation between obligatory attributes and missing attribute values \cite{Lajus:2018}.
In our previous work, we considered the problem of integration and fusion of event-centric information spread across different knowledge graphs 
and created the EventKG knowledge graph that integrates such information \cite{Gottschalk:2018, Gottschalk:2019}.
\cite{TempelmeierDD18} addressed the inference of missing categorical information in event descriptions in Web markup.
These works emphasize the need for knowledge graph completion, in particular regarding event-centric information.

\textbf{Knowledge Graph Completion}. None of the knowledge graph completion and refinement tasks has yet considered the inference of new nodes given only the knowledge graph itself \cite{wang2017knowledge, paulheim2017knowledge}. Paulheim \cite{paulheim2017knowledge} identifies three different knowledge graph completion approaches:
(i) \textit{Type Assertions Completion}. Type assertions completion is the task of predicting a type or a class of an entity \cite{paulheim2017knowledge}. A common approach to this task is to probabilistically exploit type information that is inherent in the statement properties \cite{paulheim2013type}. 
(ii) \textit{Link Prediction}. With link prediction, a ranked lists of candidates for the missing item of an incomplete triple is generated,
typically based on embeddings as performed in the TransE \cite{Bordes:2013}, STransE \cite{Nguyen:2016} and other graph embedding models \cite{Shi:2017,wang2017knowledge}.
In \approach{} we generate new events not originally present in the knowledge graph and profit from the inclusion of textual and tempo-spatial features on top of embeddings. 
(iii) \textit{External Methods}. Information extraction approaches and graph algorithms can be used to detect new relations \cite{razniewski2016but} or entity/event nodes \cite{kuzey2014fresh} from external textual data.
Instead, \approach{} solely relies on the information inherent to the knowledge graph and does not depend on the availability of the text corpora.

\textbf{Knowledge Graph Completion Tools}: A recent survey of link discovery frameworks is provided in \cite{NentwigHNR17}.
As human-curated knowledge graphs such as Wikidata demand a high quality of inserted data, there have been several tools developed that help integrating automatically generated information with the respective knowledge graph. This includes the Primary Sources Tool \cite{pellissier2016freebase}, where suggestions for new relations are confirmed by humans and \cite{Darari:2017} that provides an overview of potentially missing information. Such tools can help to integrate inferred event series data into existing knowledge graphs.

\section{Conclusion}
\label{sec:conclusion}

In this paper we addressed a novel problem of event series completion in a knowledge graph. 
The proposed \approach{} approach 
predicts sub-event relations and real-world events missing in the 
knowledge graph and does not require any external sources. 
Our evaluation on Wikidata and DBpedia datasets shows that \approach{} predicts nearly $90,000$ sub-event relations missing in Wikidata (in DBpedia: over $6,000$), clearly outperforming the embedding-based baseline by more than $50$ percentage points, and infers over $5,000$ new events (in DBpedia: over $1,300$) with a precision of $70\%$. These events and relations can be used as valuable suggestions for insertion in Wikidata and DBpedia after manual verification. 
We make our dataset publicly available to encourage further research.

\subsubsection*{Acknowledgements.} 
This work was partially funded by the EU Horizon 2020 under MSCA-ITN-2018 ``Cleopatra'' (812997), and the Federal Ministry of Education and Research, Germany (BMBF) under ``Simple-ML'' (01IS18054).

\bibliographystyle{splncs04}
\balance
\bibliography{bib_short.bib}
\end{document}